# Time-Resolved Intraband Relaxation of Strongly-Confined Electrons and Holes in Colloidal PbSe Nanocrystals

Jeffrey M. Harbold(1), Frank W. Wise(1), Hui Du(2), Todd D. Krauss(2), Kyung-Sang Cho(3), and Chris B. Murray(3)

(1) Department of Applied Physics, Cornell University, Ithaca, NY

(2) Department of Chemistry, Univ. of Rochester, Rochester, NY

(3) IBM, T. J. Watson Research Center, Yorktown Heights, NY

The relaxation of strongly-confined electrons and holes between 1P and 1S levels in colloidal PbSe nanocrystals has been time-resolved using femtosecond transient absorption spectroscopy. In contrast to II-VI and III-V semiconductor nanocrystals, both electrons and holes are strongly confined in PbSe nanocrystals. Despite the large electron and hole energy level spacings (at least 12 times the optical phonon energy), we consistently observe picosecond time-scale relaxation. Existing theories of carrier relaxation cannot account for these experimental results. Mechanisms that could possibly circumvent the phonon bottleneck in IV-VI quantum dots are discussed.

# INTRODUCTION

A dramatic reduction in the relaxation rate of carriers in three dimensionally-confined quantum systems is predicted to occur if the energy-level spacing is several times greater than the phonon energy [1, 2]. While an interesting scientific issue in nanocrystal physics, this "phonon bottleneck" would also have major implications for applications of quantum dots (QDs). In a QD gain medium, for instance, excited carriers should rapidly relax non-radiatively to their lowest states before the emission of a photon.

In prior experiments on colloidal QDs, ultrafast intraband carrier relaxation has been observed and a mechanism explaining the process has been widely adopted. This "Auger-like" electron-hole scattering process, described by Efros *et al.*, relies on the high density of hole states in II-VI and III-V semiconductor QDs plus the assumption of fast hole relaxation via phonon emission [3]. In this case, electron relaxation occurs by coupling to the large density of hole states and happens on a picosecond to sub-picosecond time scale. The strongest evidence in support of the Auger-like process comes from experiments in which this mechanism is intentionally obstructed. When the electron is forced to relax in the absence of a spectator hole, an increase of an order-of-magnitude in the relaxation time is observed [4-6]. It remains a mystery how the intraband relaxation occurs without the Auger-like mechanism.

QDs of the IV-VI semiconductors PbS, PbSe, and PbTe differ substantially from their II-VI and III-V counterparts. With large electron and hole Bohr radii (for example, in PbSe $a_e = a_h = 23$ nm), both electrons *and* holes are strongly confined. This leads to simple and sparse energy spectra for IV-VI QDs [7]. Such materials should be ideal for studying electron dynamics in a system with energy levels spaced more than a phonon-energy apart. In contrast, the electron is confined only in small II-VI and III-V QDs, and the hole is *never* strongly-confined (for example, in CdSe $a_e = 3$ nm, $a_h = 1$ nm). In addition, the confinement-induced coupling between the three (heavy-hole, light-hole, and split-off) valence bands further complicates the already dense ladder of hole states and leads to congested energy spectra in II-VI and III-V QDs. Intraband electron relaxation in II-VI and III-V nanocrystals [4-6] is dominated by processes that cannot occur in IV-VI materials, so studies of electron relaxation in II-VI and III-V nanocrystals do not provide direct information about the relevant processes in IV-VI structures.

A consequence of the sparse spectra of IV-VI QDs is that carrier relaxation *via* the Auger-like process should be impossible. Therefore, it is surprising that prior studies of IV-VI QDs did not observe a phonon bottleneck [8, 9]. Here, we report the first experiments to directly time-resolve the intraband electron and hole relaxation in strongly-confined colloidal PbSe nanocrystals [10]. At low photo-excited carrier density, the 1P to 1S relaxation time is observed to increase from 3 to 9 ps as the QD diameter increases from 4.3 to 6.0 nm. At high carrier density, the relaxation time ranges between 2 and 3 ps. In cases where carriers are pumped into higher excited states, we observe a relaxation time of ~3 ps, independent of the carrier density. No known mechanism accounts for these fast relaxations, and some possibilities are discussed.

## **EXPERIMENT**

High-quality colloidal PbSe NCs were synthesized as reported previously in the literature [11, 12]. A colloidal silica host was developed and the PbSe QDs were dispersed into the matrix [13]. The resulting solid films were 240-μm thick and contained approximately 0.1% volume occupancy of QDs. Optical characterization of the films reveals excitonic structure in the absorption, and emission spectra, as well as photoluminescence lifetimes comparable to those of the colloids (Figure 1).

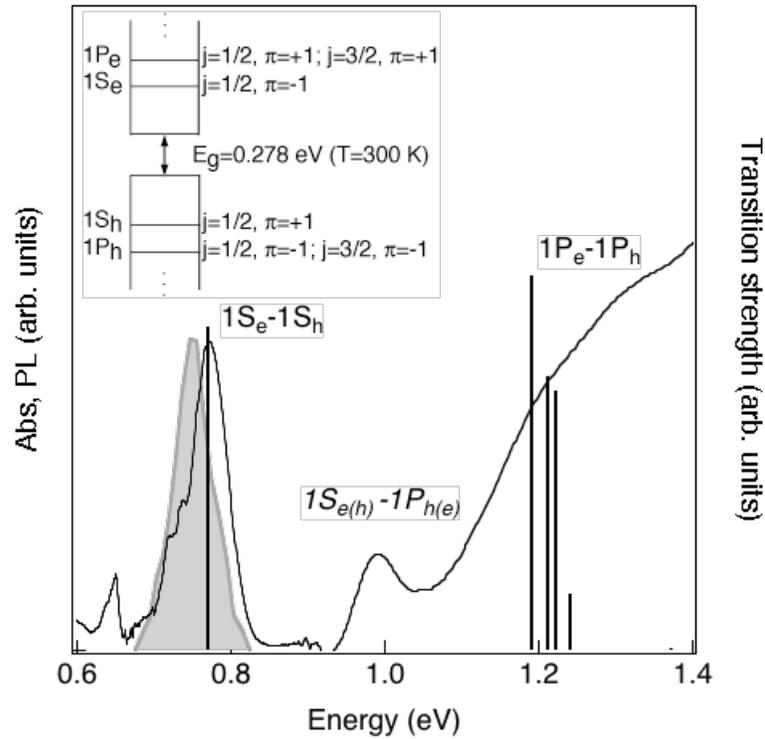

**Figure 1.** The absorption (solid line) and photoluminescence (shaded region) spectra for 6-nm PbSe QDs in the colloidal silica film. The lowest few electronic states are calculated using a 4-band envelope function formalism (inset). The strength of the lowest dipole-allowed transitions are indicated by vertical lines. The second absorption feature (italicized label) is discussed in the text. Separate measurements of a blank film that contains no QDs show that the absorption peak near 0.6 eV is due to the host.

The electronic structure of PbSe and PbS NCs was calculated by Kang *et al.* using a 4-band envelope function formalism [14]. The first 2 electron and 2 hole states and the lowest dipole-allowed transitions for a 6-nm diameter PbSe QD are shown in Figure1. The states are labeled by total angular momentum (J) and parity ($\pi$) but for simplicity, we adopt the usual nomenclature and refer to the states according to the angular momentum of the envelope function ($\ell = 0, 1\ldots$ as S, P,…). To date, electronic-structure calculations of Pb-salt QDs fail to reproduce the second absorption peak and some attention has been paid to this previously[14],[9, 15-18]. The energy of this peak matches that of the dipole-forbidden transitions between $1S_{e\,(h)}$ and $1P_{h\,(e)}$ states. However, Allan *et al*. did not find significant strength for this transition even when considering asymmetric QDs in their tight-binding calculations [18]. Still, we are intrigued to find that our ultrafast measurements support s the assignment of the second absorption peak to $1S_{e\,(h)}$ - $1P_{h\,(e)}$ transitions (see below). The nature of this unexplained transition will be more fully addressed in a future work but in the remainder of this paper, we will assume this peak arises from $1S_{e\,(h)}$ - $1P_{h\,(e)}$ transitions. For the PbSe QD sizes studied here, the $1P_{e\,(h)}$ to $1S_{e\,(h)}$ energy spacing is between 200 and 300 meV.

The vibrational modes are also modified by confinement and in QDs of polar materials, the optical mode with $\ell = 0$ is expected to contribute most strongly to relaxation by phonon emission. As shown in recent experiments by Hyun *et al*., this mode has an energy of approximately 17 meV over the range of QD sizes studied here [19]. Under these circumstances, a phonon bottleneck is expected because the $1P_{e\,(h)}$-$1S_{e\,(h)}$ energy spacing is 12 to 18 phonon energies .

Two-color transient absorption measurements were performed to investigate the intraband carrier dynamics. In this experiment, a strong pump pulse incident on the sample generates excited electron-hole pairs. A weak, time-delayed probe pulse monitors the absorption of the lowest optical transition. In the limit that $\alpha L < 0.1$ ($\alpha$ is the linear absorption coefficient and L the propagation length), the differential transmittance (DT) of the probe beam, $\Delta T/T_0 = (T-T_0)/T_0$, is directly proportional to the sum of the electron and hole populations. Here, T and $T_0$ are the probe beam transmission with and without the pump. The subsequent relaxation of the photo-excited carriers is observed by monitoring the DT as a function of the time delay between pump and probe pulses. We are unable to separate electron and hole contributions to the DT signal because of their similar electronic structures. Therefore, the mirror image transitions are always implied.

The pump and probe pulses were generated in identical optical parametric amplifiers (OPAs), simultaneously pumped by a Ti:sapphire regenerative amplifier. Each OPA was tunable in wavelength from 450 to 3000 nm. The pump and probe pulses were typically 100 fs in duration, and the time resolution of the measurement was better than 300 fs for all combinations of wavelengths employed in this work. The pump beam was mechanically chopped and the probe beam was split into signal and reference beams and measured in balanced detection using a lock-in amplifier. The probe beam was always set to the peak absorption wavelength of the lowest optical transition, $1S_e$ - $1S_h$, and two different cases were investigated. In case 1, the pump beam was tuned to the $1S_{e\,(h)}$ - $1P_{h\,(e)}$ transitions. In case 2, carriers were excited into much higher excited states by pumping with visible light ($\hbar\omega$ = 2.4 eV). The pump pulse energies were typically varied over the range of tens to hundreds of nanoJoules. All of the measurements presented here were conducted at room temperature.

## *1S - 1P Excitation*

Figure 2 shows a typical measurement when the pump photon energy is resonant with the $1S_{e\,(h)}$ - $1P_{h\,(e)}$ transition. In the left panel, we observe the population of the 1S states increasing on the time scale of the pump pulse, and then decaying within 40 to 100 ps. To focus on carriers entering 1S states, we expand the time scale in the right panel.

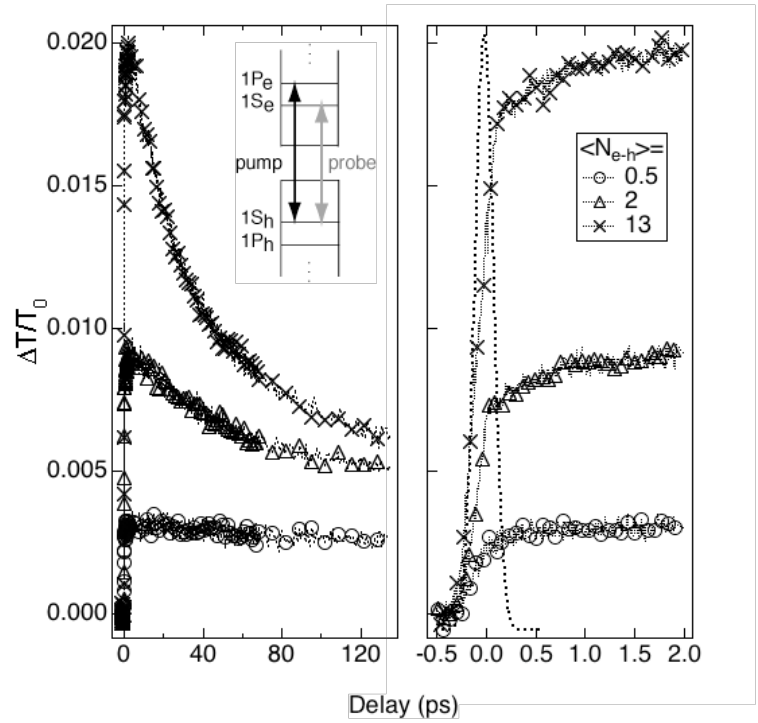

**Figure 2.** Room-temperature population dynamics of the 1S states, for the indicated densities of photo-excited carriers. The pump pulse is resonant with the $1S_h$ - $1P_e$ transition. The mirror-image $1S_e$ - $1P_h$ transition is also pumped, but for clarity this is not shown on the energy-level diagram (inset). The expanded time scale (right panel) shows carriers entering the 1S states. The dotted line is the instrument response.

The population of the 1S states exhibits two components, one within the time duration of the pump pulse and the other rising for several picoseconds thereafter. The energy-level diagram (Figure 2, inset) illustrates their origin: (1) the component rising with the pump pulse develops because the pumped and probed transitions share a state (in the case shown, $1S_h$); thus, one part of the transition is populated directly by the pump pulse, and (2) the slower rising component is caused by carriers undergoing intraband relaxation ($1P_e$ to $1S_e$ for the case shown). The first component supports our assignment of the second absorption peak to $1S_{e\,(h)}$ - $1P_{h\,(e)}$ transitions.

The magnitude of these signals also contains valuable information. We would expect the fast- and slow-rising components to contribute equally to the total signal magnitude on the basis of the similar electronic structure. However, the measurements indicate that the 1S states are almost 7-times more likely to be populated by direct excitation than by the 1P to 1S relaxation. This observation suggests the presence of an alternate decay channel. However in what follows, we will focus on the 1P to1S relaxation process.

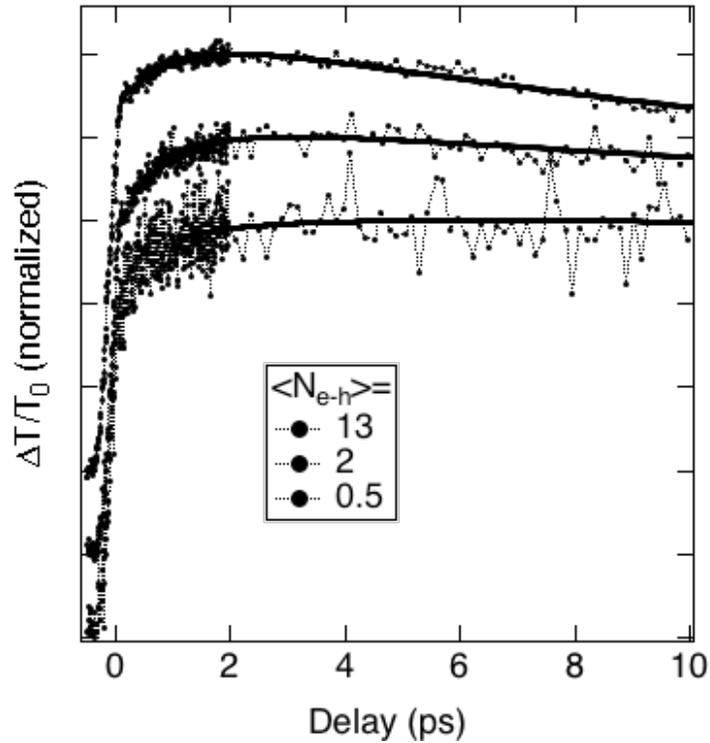

**Figure 3.** The population dynamics of the $1S_e$ and $1S_h$ states under $1S_{e\,(h)}$ - $1P_{h\,(e)}$ excitation, for different excitation levels. $<N_{e-h}>$ is the average number of photo-excited electron-hole pairs per QD. The time at which the population peaks is extracted from a fit to the data (solid lines). The traces are normalized and then offset for clarity. The delay stage step size was changed at 2 ps, giving the appearance of an abrupt change in noise level.

Figure 3 shows the DT traces for the sample in Figure 2 over a wider ranger of carrier densities, but normalized to the peak DT. The curves are fit to extract the time at which the $1S_e$ - $1S_h$ population peaks, and this is taken to be the 1P to1S relaxation time. With an average of 0.5 electron-hole pairs per QD

($<N_{e-h}> = 0.5$), the relaxation is observed to occur in 6 ps, and this time decreases as the photo-excited carrier density is increased. We observe this trend in all the QDs studied. The fastest relaxation time observed is 2 ps, which is approximately 3 – 4 times larger than the electron relaxation time in bulk PbSe. We repeated these measurements on a range of QD sizes between 4.3 and 6.0 nm and find the 1P to 1S relaxation time increases with QD size (Figure 4). In these measurements the carrier density was kept low to avoid multi-carrier effects in the relaxation.

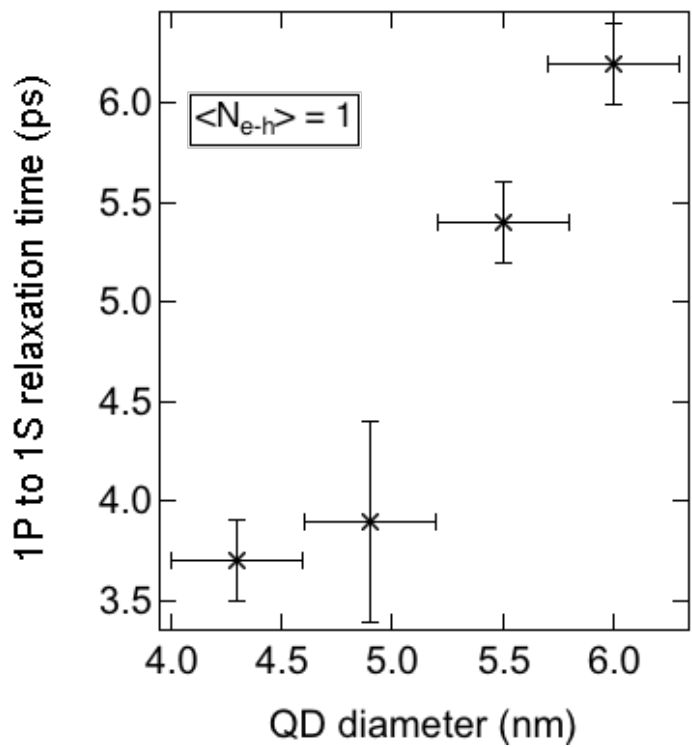

**Figure 4.** 1P - 1S relaxation time is plotted versus QD diameter.

### *High-energy excitation*

We also explored the effect of exciting carriers far above the 1P states. This has the advantage of not directly populating a state that is probed. Accordingly, the rise of the $1S_e$ - $1S_h$ population shown in figure 5 now consists of a single component. When a 4.9-nm-diameter QD is pumped at 2.4 eV, we find that the $1S_e$ - $1S_h$ population develops in approximately 3 ps at both low ($<N_{e-h}>$ <1 ) and high ($<N_{e-h}>$ = 10) carrier densities. This insensitivity to carrier density is in contrast to the results obtained when exciting directly into 1P states. Therefore, we need to consider these two regimes separately.

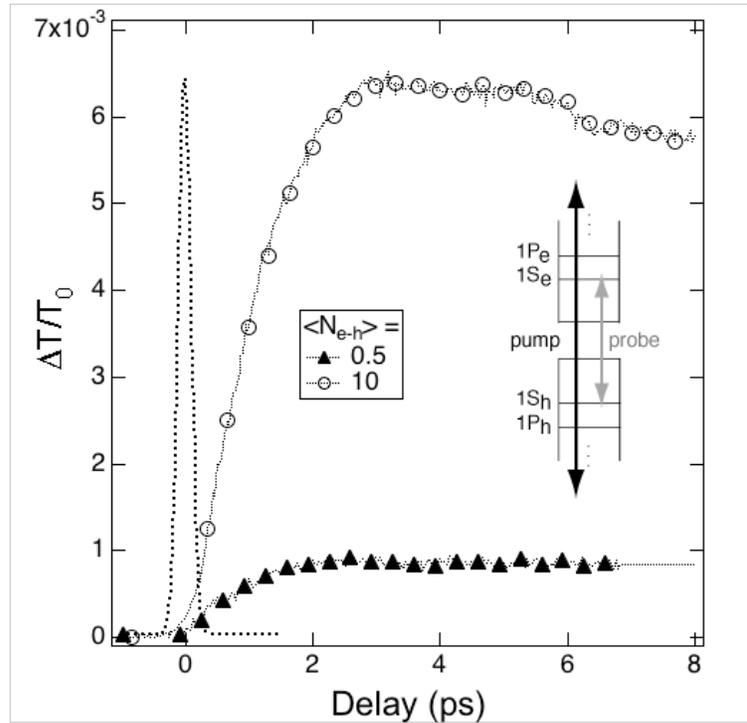

**Figure 5.** Population dynamics of the $1S_e$ and $1S_h$ states for a 2.4-eV pump and the indicated photo-excited carrier densities.

## DISCUSSION

We now turn to a discussion of the possible relaxation mechanisms. One way to explain the intraband relaxation of carriers from states far above 1P (Figure 5) and the observed insensitivity to carrier density is impact ionization. In this process, a carrier with energy in excess of twice the $1S_e$ - $1S_h$ energy difference ($\Delta E_{1S-1S}$) relaxes to its 1S state by exciting additional electrons from $1S_h$ to $1S_e$ [20]. Thus, two or more cooled carriers may result from a single highly-excited one. Schaller *et al*. have attributed the appearance of multi-exciton signatures in the decay of the 1S states with $<N_{e-h}>$ < 1 to impact ionization[21]. At low carrier density and high-energy excitation ($\hbar\omega = 2.5*\Delta E_{1S-1S}$), we also observe a fast decay component from the 1S states. We do not observe this for 1S – 1P excitation.

The 1P – 1S relaxation requires a different explanation. Any possible role of multi-carrier effects is excluded by consideration of the relaxation for $<N_{e-h}> < 1$. Yet even in this case, the observed relaxation time is between 3 and 9 ps, with the smallest QDs exhibiting faster relaxation. These results are in reasonable agreement with the upper bound on the intraband relaxation time set by Wehrenberg *et al.* [9].

The contribution of multiphonon [22] or polaron [23] effects in the relaxation are ruled out because they cannot account for efficient relaxation over such large energy level spacings. However, there may be additional states that could facilitate the relaxation. One possible source of such states is the fact that there are eight equivalent L-valleys in PbSe. Splitting of these degenerate states is calculated to be as large as ~ 100 meV in the smallest QDs [18]. Even with such splitting, however, the energy gap between the 1P and 1S manifolds would be several phonon energies [24].

It is also possible that the surface could play a role in the relaxation we observe. Darugar *et al.* have studied the relaxation from the lowest transition in CdSe QDs and quantum rods and find that QDs relax faster than quantum rods of the same diameter. Darugar *et al.* correlate this to the predicted change in excited electron density on the nanostructure surface, and conjecture that coupling between the excited electrons and large surface molecules could lead to efficient carrier thermalization [25] .

Before concluding, we note the similarity in intraband relaxation times (1) reported in this work for electrons and holes in PbSe QDs and (2) observed for electrons in II-VI and III-V semiconductor QDs when the Auger-like electron-hole scattering mechanism is intentionally obstructed [4-6]. These results suggest the existence of a non-radiative relaxation channel on the 1-10 ps time scale that is accessible to strongly-confined electrons and/or holes in colloidal QDs, *independent of material system*. This too suggests that the surface may be an integral part of the intraband relaxation mechanism. Renewed efforts to incorporate the surface's role into the relaxation may be needed .

**CONCLUSIONS**

In conclusion, we have time-resolved the relaxation of electrons and holes in colloidal PbSe NCs. In stark contrast to other semiconductor NCs, both electrons and holes are strongly confined in PbSe QDs. Therefore, the Auger-like electron-hole scattering cannot be invoked to explain the intraband relaxation of charge carriers. In cases where carriers are pumped into higher excited states, we observe a relaxation time of ~3 ps, independent of the carrier density. We tentatively attribute this behavior to carrier cooling *via* impact ionization. When the P states are directly populated with, on average, less than one electron or hole per QD, the 1P to 1S relaxation time increases from 3 to 9 ps as the QD size is varied between 4.3 and 6.0 nm in diameter. The size dependence of this relaxation may be correlated with the change in electron density on QD surface as previously suggested in the literature. This could explain the similarity of the relaxation times observed in PbSe QDs and in surface-modified colloidal II-VI and III-V QDs. Known mechanisms based on interior electron states and vibrational modes cannot account for the experimental results, which suggests that surface atoms and ligands may play a major role.

This work was primarily supported by the Nanoscale Science and Engineering Initiative of the National Science Foundation under NSF Award Number EEC-0117770. The laser facility used for these experiments was supported under NIH award 9 P41 EB001976-17 and NSF award CHE-0242328. We also thank J. Gillies and D. Landry of Evident Technologies, Inc. (Troy, NY) for the QD films.